\documentclass[11pt,twoside]{stm}

\begin{document}
\markboth{\sc E.~Jur\v{c}i\v{s}inov\'a, M.~Jur\v{c}i\v{s}in}{\sc
Scaling Regimes of a Passive Scalar Field}

\STM

\title{Combined effects of compressibility and helicity on the scaling regimes of a
passive scalar advected by turbulent velocity field with finite
correlation time}

\authors{E.~Jur\v{c}i\v{s}inov\'a$^{1,2}$, M.~Jur\v{c}i\v{s}in$^{1,3}$}

\address{$^{1}$\,IEP SAS, 040 01 Ko\v{s}ice, Slovakia,\\$^{2}$\,LIT JINR, 141 980 Dubna, Russia,
\\$^{3}$\,BLTP JINR, 141 980 Dubna, Russia}
\bigskip

\begin{abstract}
The influence of compressibility and helicity on the stability of
the scaling regimes of a passive scalar advected by a Gaussian
velocity field with finite correlation time is investigated by the
field theoretic renormalization group within two-loop approximation.
The influence of helicity and compressibility on the scaling regimes
is discussed as a function of the exponents $\varepsilon$ and
$\eta$, where $\varepsilon$ characterizes the energy spectrum of the
velocity field in the inertial range $E\propto k^{1-2\varepsilon}$,
and $\eta$ is related to the correlation time at the wave number $k$
which is scaled as $k^{-2+\eta}$. The restrictions given by nonzero
compressibility and helicity on the regions with stable infrared
fixed points which correspond to the stable infrared scaling regimes
are discussed. A special attention is paid to the case of so-called
frozen velocity field when the velocity correlator is time
independent. In this case, explicit inequalities which must be
fulfilled in the plane $\varepsilon-\eta$ are determined within
two-loop approximation.
\end{abstract}

\section*{Introduction}

One of the main problems in the modern theory of fully developed
turbulence is to verify the validity of the basic principles of
Kolmogorov-Obukhov (KO) phenomenological theory and their
consequences within the framework of a microscopic model
\cite{MonYag75,Frisch95}. On the other hand, recent experimental,
numerical and theoretical studies signify the existence of
deviations from the well-known Kolmogorov scaling behavior. The
scaling behavior  of the velocity fluctuations with exponents, which
values are different from Kolmogorov ones, is known as anomalous and
is associated with intermittency phenomenon \cite{Frisch95}. Even
thought the understanding of the intermittency and anomalous scaling
within the theoretical description of the fluid turbulence on basis
of the "first principles", i.e., on the stochastic Navier-Stokes
equation, still remains an open problem, considerable progress has
been achieved in the studies of the simplified model systems which
share some important properties of the real turbulence.

The crucial role in these studies is played by models of advected
passive scalar field \cite{Obu49}. Maybe the most known model of
this type is a simple model of a passive scalar quantity advected by
a random Gaussian velocity field, white in time and self-similar in
space, the so-called Kraichnan's rapid-change model \cite{Kra68}. It
was shown by both natural and numerical experimental investigations
that the deviations from the predictions of the classical KO
phenomenological theory is even more strongly displayed for a
passively advected scalar field than for the velocity field itself
(see, e.g., \cite{FaGaVe01} and references cited therein). At the
same time, the problem of passive advection is much more easier to
be consider from theoretical point of view. There, for the first
time, the anomalous scaling was established on the basis of a
microscopic model \cite{Kraichnan94}, and corresponding anomalous
exponents was calculated within controlled approximations (see
review \cite{FaGaVe01} and references therein).

In paper \cite{AdAnVa98+} the field theoretic renormalization group
(RG) and operator-product expansion (OPE) were used in the
systematic investigation of the rapid-change model. It was shown
that within the field theoretic approach the anomalous scaling is
related to the very existence of so-called "dangerous" composite
operators with negative critical dimensions in OPE (see, e.g.,
\cite{Vasiliev,AdAnVa99} for details).

Afterwards, various generalized descendants of the Kraichnan model,
namely, models with inclusion of large and small scale anisotropy
\cite{AdAnHnNo00}, compressibility \cite{AdAn98} and finite
correlation time of the velocity field \cite{Antonov99,Antonov00}
were studied by the field theoretic approach. General conclusion is:
the anomalous scaling, which is the most important feature of the
Kraichnan rapid change model, remains valid for all generalized
models.

In paper \cite{Antonov99} the problem of a passive scalar advected
by the Gaussian self-similar velocity field with finite correlation
time \cite{all2} was studied by the field theoretic RG method.
There, the systematic study of the possible scaling regimes and
anomalous behavior was present at one-loop level. The two-loop
corrections to the anomalous exponents were obtained in
\cite{AdAnHo02}. In paper \cite{Antonov00} the influence of
compressibility on the problem studied in \cite{Antonov99} was
analyzed. The effects of the presence of helicity in the system on
the scaling regimes and anomalous dimensions were studied in
\cite{ChHnJuJuMaRe061,ChHnJuJuMaRe06} within of two-loop
approximation. It was shown that although the separate composite
operators which define anomalous dimensions strongly depend on the
helicity parameter the resulting two-loop contributions to the
critical dimensions of the structure functions are independent of
helicity. This rather intriguing result can have at least two
independent explanations. First, it can be simply a two-loop result
which will be changed when one will study three-loop approximation.
Second, more interesting conclusion (but it is only a speculation
for now) is that this situation will be held at each level of the
perturbation theory while the assumption of incompressibility or
isotropy will be supposed. From this point of view, the
investigation of the compressible system together with assumption of
helicity of the system within two-loop approximation can give an
interesting answer. For this purpose in \cite{HnJuJuRe06} the
influence of compressibility of the model on the stability of the
scaling regimes was studied and the restrictions on the parameter
space were analyzed. In what follows, we shall continue these
studies and our aim will be to find combine effects of helicity and
compressibility on the stability of the scaling regimes in two-loop
approximation. It can be consider as the starting point for the
subsequent investigation of the anomalous scaling of the correlator
or structure functions of a passive scalar.

\section*{Formulation of the model}

The advection of a passive scalar field is described by equation
\begin{equation}
\partial_t \theta + ({\bf v} \cdot {\bf \partial})\theta =\nu_0
\nabla \theta + f^{\theta}\,, \label{model}
\end{equation}
where $\theta(x)=\theta(t,{\bf x})$ is a passive scalar field,
$\nu_0$ is the molecular diffusivity coefficient, ${\bf v}(x)$ is
compressible velocity field, and $f^{\theta} \equiv f^{\theta}(x)$
is a Gaussian random noise with zero mean and correlation function
\begin{equation}
\langle f^{\theta}(x) f^{\theta}(x^{\prime})\rangle =
\delta(t-t^{\prime})C({\bf r}/\tilde{L}), \,\,\, {\bf r}={\bf
x}-{\bf x^{\prime}},\label{correlator}
\end{equation}
where parentheses $\langle...\rangle$ denote average over
corresponding statistical ensemble. The noise maintains the
steady-state of the system but its concrete form is not essential.
$\tilde{L}$ denotes an integral scale related to the stirring.

In real problems the velocity field ${\bf v}(x)$ satisfies
stochastic Navier-Stokes equation but we shall suppose that the
velocity field obeys a Gaussian distribution with zero mean and
correlator
\begin{equation}
\langle v_i(x) v_j(x^{\prime})\rangle=\int
\frac{d\omega}{2\pi}\int \frac{d{\bf k}}{(2\pi)^d} R_{ij}({\bf k})
D_v(\omega,k) e^{-i\omega(t-t^{\prime})+i{\bf k}\cdot ({\bf
x}-{\bf x^{\prime}})}\,.\label{model1}
\end{equation}
Here $R_{ij}({\bf k})=P_{ij}({\bf k}) + Q_{ij}({\bf k}) +
H_{ij}({\bf k})$, where $P_{ij}({\bf k})=\delta_{ij}-k_i k_j/k^2$ is
the non-helical transverse projector, $Q_{ij}({\bf k})=\alpha k_i
k_j/k^2$ is longitudinal projector with the compressibility
parameter $\alpha\geq 0$, and $H_{ij}({\bf k})= i \rho
\varepsilon_{ijl} k_l/k$ is the helical transverse projector with
helicity parameter $\rho\in \langle 0,1 \rangle$
($\varepsilon_{ijl}$ is Levi-Civita's completely antisymmetric
tensor of rank 3, and $k\equiv |{\bf k}|$). $d$ is the
dimensionality of the ${\bf x}$ space which must be taken $d=3$ in
the helical case ($\rho>0$). The function $D_v$ is taken in the form
\begin{equation}
D_v(\omega,k)=\frac{g_0 \nu_0^3
k^{4-d-2\varepsilon-\eta}}{\omega^2+(u_0 \nu_0 k^{2-\eta})^2}\,,
\label{model15}
\end{equation}
where $g_0>0$ and the exponent $\varepsilon$ describe the equal-time
velocity correlator (the energy spectrum). On the other hand, the
constant $u_0$ and the exponent $\eta$ are related to the frequency
$\omega$, characteristic of the mode $k$. In our model the exponents
$\varepsilon$ and $\eta$ play the role of the RG expansion
parameters. The model contains two special cases. In the limit
$u_0\rightarrow\infty$ and $g_0/u^2_0=const.$ one obtains the
so-called rapid-change model, and the limit $u_0\rightarrow 0$ and
$g_0/u_0=const.$ corresponds to the case of a "frozen" velocity
field \cite{Antonov99,Antonov00}.

The stochastic problem (\ref{model})-(\ref{model1}) is equivalent to
the field theoretic model of the set of fields $\Phi \equiv
\{\theta, \theta^{\prime}, {\bf v}\}$ (see, e.g., \cite{Vasiliev})
with action functional
\begin{eqnarray}
\hspace{-1cm} S(\Phi)=&-&\frac{1}{2} \int dt_1\,d^d{\bf
x_1}\,dt_2\,d^d{\bf x_2} \,\,v_i(t_1,{\bf x_1}) [D^{v}_{ij}(t_1,{\bf
x_1};t_2,{\bf x_2})]^{-1} v_j(t_2,{\bf x_2})  \nonumber \\
&+& \int dt\,d^d{\bf x}\,\, \theta^{\prime}\left[-\partial_t \theta
- v_i\partial_i\theta+\nu_0\triangle\theta \right], \label{action1}
\end{eqnarray}
where the unimportant term related to the noise (\ref{correlator})
is omitted, $\theta^{\prime}$  is an auxiliary scalar field, and
summations are implied over the vector indices. The second line in
(\ref{action1}) represents the Martin-Siggia-Rose action for the
stochastic problem (\ref{model}) at fixed velocity field ${\bf v}$,
and the first line describes the Gaussian averaging over ${\bf v}$
defined by the correlator $D^v$ in (\ref{model1}) and
(\ref{model15}).


The functional formulation (\ref{action1}) means that statistical
averages of random quantities in the stochastic problem defined by
(\ref{model}) and (\ref{model1}) corresponds to functional averages
with the weight $S(\Phi)$.



\section*{UV renormalization, fixed points and scaling
regimes}

Using the standard analysis of quantum field theory one can find
that the UV divergences are only present in the
one-particle-irreducible Green functions $\langle \theta^{\prime}
\theta \rangle_{1-ir}$ and $\langle\theta^{\prime} \theta {\bf
v}\rangle_{1-ir}$.  The renormalized action functional has the
following form
\begin{eqnarray}
S(\Phi)=&-&\frac{1}{2} \int dt_1\,d^d{\bf x_1}\,dt_2\,d^d{\bf x_2}
v_i(t_1,{\bf x_1}) [D_{ij}^v(t_1,{\bf
x_1};t_2,{\bf x_2})]^{-1} v_j(t_2,{\bf x_2}) \label{actionRen} \\
&+& \int dt\,d^d{\bf x}\,\, \theta^{\prime}\left[-\partial_t \theta
- Z_2 v_i\partial_i\theta+\nu Z_1 \triangle\theta \right],\nonumber
\end{eqnarray}
The dimensionless parameters $g,u$ and $\nu$ are the renormalized
counterparts of the bare parameters (denoted by $0$). $Z_1$ and
$Z_2$ are the renormalization constants of our model. We shall
calculate them in 2-loop approximation in the minimal subtraction
scheme. Our model is multiplicatively renormalizable which is
represented by the following relations between renormalized and bare
parameters (see, e.g., \cite{Antonov99,Antonov00}):
\begin{equation}
\nu_0=\nu Z_{\nu},\,\,\, g_0=g \mu^{2\varepsilon+\eta}
Z_g,\,\,\,u_0=u\mu^{\eta} Z_u,\,\,\, {\bf v}\rightarrow Z_v {\bf v}.
\label{zetka}
\end{equation}
where
\begin{equation}
Z_{\nu}=Z_1,\,\,\,Z_u=Z_1^{-1},\,\,\,Z_g=Z_2^2 Z_1^{-3},\,\,\,
Z_v=Z_2. \label{zetka1}
\end{equation}
and $\mu$ is the renormalization mass. One and two-loop Feynman
diagrams which contribute to the renormalization constants $Z_1$ and
$Z_2$ can be found in \cite{HnJuJuRe06}. They have the following
form
\begin{equation}
Z_1=\frac{g}{\varepsilon} A_1 +
\frac{g^2}{\varepsilon}\left(\frac{1}{\varepsilon}B_1+C_1\right),
\quad  Z_2=\frac{g}{\varepsilon} A_2 +
\frac{g^2}{\varepsilon}\left(\frac{1}{\varepsilon}B_2+C_2\right),
\label{const}
\end{equation}
where the one-loop contributions are (in the MS-scheme)
\begin{equation}
A_1= -\frac{S_d}{(2\pi)^d}\frac{1}{4 u (1+u)^2}\frac{(1 + u)(d - 1 +
\alpha) - 2\alpha}{d}, \quad A_2=\frac{S_d}{(2\pi)^d}\frac{\alpha}{4
u (1+u)^2},
\end{equation}
where $S_d$ is $d$ dimensional sphere given by the expression $S_d=2
\pi^{d/2}/\Gamma(d/2)$. Two-loop contributions will not be shown
explicitly here (the explicit form of the corresponding expressions
in the non-helical case can be found in \cite{HnJuJuRe06}, and the
explicit form of the helical two-loop contribution to the $Z_1$ can
be found in \cite{ChHnJuJuMaRe061}).

From the renormalization constants (\ref{const}) one can define the
corresponding anomalous dimensions $\gamma_i=\mu \partial_{\mu} \ln
Z_i$ and $\beta$ functions for all invariant charges ($X=g,u$):
$\beta_X=\mu\partial_{\mu} X$. In our case we have
\begin{equation}
\gamma_1\equiv \mu \partial_{\mu} \ln Z_1 = -2 (g A_1 + 2 g^2
C_1),\quad \gamma_2\equiv \mu \partial_{\mu} \ln Z_2 = -2 (g A_2 + 2
g^2 C_2),\label{gammas}
\end{equation}
and
\begin{equation}
\beta_g \equiv \mu \partial_{\mu} g =g
(-2\varepsilon-\eta+3\gamma_1-2\gamma_2),\quad \beta_u \equiv \mu
\partial_{\mu} u = u(-\eta+\gamma_1).\label{betas}
\end{equation}

Now one has all necessary tools to begin with analysis of possible
scaling regimes of the model. It is well known that possible scaling
regimes of a renormalizable model are associated with the infrared
(IR) stable fixed points of the corresponding RG equations. The
fixed points are determined from the requirement that all $\beta$
functions are vanish: $\beta_X(X_*)=0$, where $X=g,u$ and $X_*$ are
coordinates of the corresponding fixed point. The type of the fixed
point is given by the matrix of the first derivatives
$\Omega_{ij}=\partial \beta_i/\partial X_j$. The fixed point is IR
stable if all the eigenvalues of the matrix $\Omega$ are positive
(precisely their real parts). In our model we have five types of
possible scaling regimes related to fixed points of the model.

First of all, we shall investigate the rapid-change limit:
$u\rightarrow\infty$. In this regime, it is necessary to make
transformation to new variables, namely, $w\equiv1/u$, and
$g^{\prime}\equiv g/u^2$, with the corresponding changes in the
$\beta$ functions:
\begin{equation}
\beta_{g^{\prime}}=g^{\prime}(\eta-2\varepsilon
+\gamma_{1}-2\gamma_2), \quad \beta_w=w(\eta-\gamma_{1}).
\label{betas1}
\end{equation}
In this case we have two fixed points which we denote as I and II.
First of them is trivial one, namely

{\bf Fixed Point I:} $w_*=1/u_*=g_*^{\prime}=0$ \\
with $\gamma_{1}=0$. The matrix $\Omega$ is diagonal with the
elements (eigenvalues): $\Omega_1=\eta$,
$\Omega_2=\eta-2\varepsilon$. Thus, the corresponding scaling regime
is IR stable if $\eta>0$ and, at the same time, $\eta>2\varepsilon$.

The second one is defined as follows

{\bf Fixed Point II:} $w_*=1/u_*=0$, and
\begin{equation}
\bar{g}_*^{\prime}=\frac{2d}{d-1+\alpha}(2\varepsilon-\eta),
\end{equation}
where we denote $\bar{g}^{\prime}=g^{\prime} S_d/(2\pi)^d$. The
anomalous dimension $\gamma_{1}$ at the fixed point is
$\gamma^*_1=2\varepsilon -\eta$. These are exact one-loop results.
It is the consequence of the non-existence of the higher-loop
corrections. Corresponding matrix of the first derivatives is
triangular with diagonal elements (eigenvalues)
$\Omega_1=2(\eta-\varepsilon)$ and $\Omega_2=2\varepsilon-\eta$.
Thus, the conditions $g_*^{\prime}>0$ and $\Omega_{1,2}>0$ for the
IR stable regime lead to the inequalities: $\eta > \varepsilon$ and
$\eta < 2\varepsilon$.

The second nontrivial limit of our general model is so-called frozen
limit given by $u \rightarrow 0$. In this case, it is again
necessary to make transformation to new variables, namely,
$g^{\prime\prime}\equiv g/u$ and $u$ is unchanged. It leads to the
changes in the $\beta$ functions which are now define as follows
($\beta_u$ is the same as in the general case)
\begin{equation}
\beta_{g^{\prime\prime}}=g^{\prime\prime}(-2\varepsilon
+2\gamma_{1}-2\gamma_2), \quad \beta_u=u(-\eta+\gamma_{1}).
\label{betas2}
\end{equation}
In this case we have again two fixed points which we denote as III
and IV. First of them is trivial one, namely

{\bf Fixed Point III:} $u_*=g_*^{\prime\prime}=0$. \\
The corresponding matrix $\Omega$ is diagonal with the elements:
$\Omega_1=-2\varepsilon$ and $\Omega_2=-\eta$. It means that the
corresponding scaling regime is IR stable if $\varepsilon<0$ and
$\eta<0$.

The second one is defined as follows

{\bf Fixed Point IV:} $u_*=0$, and
\begin{equation}
\bar{g}_*^{\prime\prime}=-\frac{\varepsilon}{2(A^{\prime\prime}_{10}-A^{\prime\prime}_{20})}-
\frac{C^{\prime\prime}_{10}-C^{\prime\prime}_{20}}{2
(A^{\prime\prime}_{10}-A^{\prime\prime}_{20})^3} \varepsilon^2,
\end{equation}
where $A^{\prime\prime}_{i0}$ and $C^{\prime\prime}_{i0}$ for
$i=1,2$ are expressions from (\ref{const}) taken in the given limit
and again $\bar{g}^{\prime\prime}=g^{\prime\prime} S_d/(2\pi)^d$. In
this case we have the relation $\gamma_1^*=\gamma_2^*+\varepsilon$
between anomalous dimensions. The matrix of the first derivatives is
triangular with eigenvalues (diagonal elements):
\begin{equation}
\Omega_1=-\eta+\gamma_1^*,\quad \Omega_2=2
g^{\prime\prime}_*\left(\frac{\partial \gamma_1}{\partial
g^{\prime\prime}}-\frac{\partial \gamma_2}{\partial
g^{\prime\prime}}\right)_*.
\end{equation}
In the helical case, when helicity parameter $\rho>0$, one must work
with $d=3$, and we obtain
\begin{eqnarray}
\bar{g}^{\prime\prime}_*&=&3\varepsilon+\varepsilon^2\left(\frac32-\frac74
\alpha-\frac{9}{32}\pi^2\rho^2\right),\\
\Omega_1 &=&\varepsilon-\eta-\frac{\varepsilon
\alpha}{2}+\frac{\varepsilon^2 \alpha}{192}(-64+56\alpha
+9\pi^2\rho^2), \quad \Omega_2 =\varepsilon\left(2+\varepsilon
\left(-1+\frac{7}{6} \alpha+\frac{3}{16} \pi^2 \rho^2
\right)\right),
\end{eqnarray}
therefore, the conditions to have $\bar{g}^{\prime\prime}_*>0$ and
$\Omega_{1,2}>0$ are
\begin{equation}
\varepsilon >0,\quad \varepsilon\left(-1+\frac{7}{6}
\alpha+\frac{3}{16} \pi^2 \rho^2 \right) > -2,\quad
\eta<\varepsilon-\frac{\varepsilon \alpha}{2}+\frac{\varepsilon^2
\alpha}{192}(-64+56\alpha +9\pi^2\rho^2).
\end{equation}
These inequalities define the region in the parameter space for
which the scaling regime is IR stable.

The last but the most interesting scaling regime is obtained when
one assume that $0 <u_* < \infty$. Let us briefly discuss this case.
The corresponding IR fixed point will be denoted as V. The
coordinates of the fixed point is now defined by the requirement of
vanishing of the $\beta$ functions which are given in (\ref{betas}).
The fixed point value for $g$ is given as

{\bf Fixed Point V:} $u_*>0$ \\
\begin{equation}
\mathrm{FPV}:\,\,\,\, g_*=-\frac{\varepsilon}{2(A_{1}-A_{2})}-
\frac{C_{1}-C_{2}}{2 (A_{1}-A_{2})^3} \varepsilon^2,
\end{equation}
where the functions $A_1, A_2, C_1$, and $C_2$ are given in
(\ref{const}), and where the parameter $u$ is taken at its fixed
point value $u_*$ which is defined implicitly by the equation
\begin{equation}
-\eta+\gamma_1^*(u_*)=0.
\end{equation}
The analysis of the problem is relatively simple in one-loop level,
where explicit expressions can be found \cite{Antonov00}. On the
other hand, as was briefly discussed in \cite{HnJuJuRe06}, the
situation is essentially more complicated when we are working in
two-loop approximation. Detail analysis shows that the investigation
of the IR stability of the fixed point in the general case of the
present model has to be done individually for concrete situation
which is rather cumbersome and it must be done in a separate work.

\section*{Conclusions}
In present paper we have studied the influence of compressibility
and helicity of the system on the possible IR scaling regimes of the
model of a passive scalar advected by a Gaussian velocity field with
finite time correlations by means of the field theoretic RG
technique. The dependence of the fixed points, which are directly
related to the existence of possible IR scaling regimes, on the
parameters of compressibility and helicity as well as their IR
stability is discussed. The explicit inequalities, which define the
stable IR scaling regimes, are found in the case of the frozen limit
of the model. The most general case with finite time correlations of
the velocity field is more complicated within two-loop approximation
and has to be consider in detail once more.

\bigskip

\noindent  ACKNOWLEDGEMENTS --- It is a pleasure to thank the
Organizing Committee of the STM-2005 for kind hospitality. The work
was supported in part by VEGA grant 6193 of Slovak Academy of
Sciences, and by Science and Technology Assistance Agency under
contract No. APVT-51-027904.

\end{document}